# The Nonlinear Impact of Minimum Wage on Labor Employment in China

Junhan LYU[1*], Tianle Zhai[2], Zicheng Peng[2], Xuhang Huang[2]

[1*] Department of Accountancy, Economics and Finance, School of Business, Hong Kong Baptist University, Shaw Campus, Hong Kong Baptist University, 34 Renfrew Road, Kowloon Tong, Kowloon, Hong Kong,

[2] School of Business, Macau University of Science and Technology, Macau, 999078, China.

*Corresponding author(s). E-mail(s): 23213078@life.hkbu.edu.hk
Contributing author: 23213078@life.hkbu.edu.hk;1220017324@student.must.edu.mo; 1220012523@student.must.edu.mo; 1220017001@student.must.edu.mo

## Abstract

This paper examines the impact of increasing minimum wages, focusing primarily on their effect on employment. Our research involved analyzing the statistics of panel data, testing fixed effects and stationary, conducting linear regression, and integrating the linear regression model with nonlinear model analysis. The results indicate that fluctuations in the employment rate are almost entirely explained by the selected explanatory variables, and there is a significant negative correlation between minimum wages and the employment rate. This paper contributes to current research by providing more comprehensive analyses, particularly through the use of nonlinear models, resulting in better-fitting models. We employed multiple fitting methods for time series data and their differentials, combining these results with nonlinear analysis.

## 1 Introduction

### 1.1 Literature review

Minimum wage policies, which are in a process of continuous adjustment in China, have persistent and vital influences on employment. Local employers are affected by these policies, and since the effects are heterogeneous among different provinces, they also impact the regional distribution of employment. The minimum wage in China varies significantly across provinces, ranging from 1650 RMB to 2590 RMB in 2023. Moreover, different provinces adjust the minimum requirements at varying frequencies (Koty and Zhou, 2020). The timing of these changes is also noteworthy. Since China became a member state of the WTO in 2001, foreign direct investment has become a vital component of China's economy. Foreign investors were primarily attracted by the relatively low wages in China (Liu and Pearson, 2010; Zhang, 2005). As a result, increases in the minimum wage affect foreign direct investment and GDP, which, in turn, influence employment. As of now, low labor costs are no longer a significant comparative advantage for China, especially when compared with neighboring states like Vietnam and India (Lewis, 2016; Wye and Bahri, 2021; Schmillen et al., 2023; Li and Lin, 2020). Generally, the effect of the minimum wage is significant and has become more pronounced with its increasing value and regional heterogeneity in recent years.

Numerous studies have been conducted using various methods to examine the relationship between minimum wage restrictions and employment in China. A basic model in labor economics is the search-matching model. Liu (2013) uses the search-matching model as a theoretical method and linear regression as empirical evidence to analyze the labor supply–demand relationship. Zheng et al. (2021) attribute mismatching as the main reason for overeducation and the wage penalty. Researching labor issues under a general equilibrium benchmark is also widely accepted. Li et al. (2020) established a small open general equilibrium model to investigate the effects of unemployment and wage inequality. They also incorporated motional simulation methods with static analysis. Zhao and Sun (2021) apply some new Keynesian assumptions, such as frictions and dynamic incentive contracts, into their model, finding that a higher minimum wage has two impacts: increasing the unemployment rate and causing a spillover effect on higher-income workers. Developing a general equilibrium model with heterogeneous firms in terms of capital intensity and industry level to predict the effect of minimum wages is also a recent area of focus (Bai et al., 2021).

Recently, researchers have focused more on regional, operational, and property heterogeneity in data and regression methods. Yang and Gunderson (2020) conducted a causal estimation of employment, hours, and wages among Chinese immigrant laborers using difference-in-differences (DID) analysis and tested the results through robustness checks. Beladi et al. (2020) use the average wage of employees at the firm level as a proxy for labor costs to determine the impact on heterogeneous companies and regions. Chu et al. (2021) examine the differing impacts on domestic and import companies using city-level panel data. Wang et al. (2023) employ an instrumental variables (IV) strategy and DID analysis to research firms' reactions to higher minimum wages. Hau et al. (2021) analyze regional variations in firms' reactions, focusing on input substitution and wage shares when experiencing minimum wage shocks.

It is worth noting that until recent years, as mentioned above, contemporary research has mainly focused on linear regression models. Although some very recent studies explore nonlinear relationships involving financial development and innovations with employment (Chen et al., 2021; Law and Law, 2024), the models they use are still in linear formats. However, the relationship between labor employment and minimum wage is typically not linear. The basic search-matching model can deduce a nonlinear relationship, and some important studies also support this (Card, 1993; Kaufman, 2020). Given this situation, it is significant to seek a nonlinear model to analyze the impact of minimum wages on employment rates.

There are many classical nonlinear models in physics and engineering. Although nonlinear models are not common in economics, we can extract methods from similar fields. Typically, we can develop a nonlinear model by analyzing the residuals of a linear model and incorporating nonlinear and linear components to construct the model. We can also study the equations governing the differentiation of variables to construct the model. Based on this, we combined the traditional OLS model in econometrics with nonlinear models to study this effect. We also analyzed the regional and time distributions of wages and a series of auxiliary variables, such as GDP per capita, living standards, and others. First, we applied OLS analysis and conducted tests on significance, correlation, and stationarity. Subsequently, we extended the research by applying a nonlinear model. The results of the two types of analysis were aligned.

Our work contributes to current research mainly by combining the two types of models and applying a novel model in labor economics research. The results show that our model provides a better fit for the regression on the relationship between minimum wages and employment in the labor market.

### 1.2 Research method and content

#### 1.2.1 Research content

This paper takes the impact of the minimum wage on labor employment as the main research objective, and studies and discusses the impact of the minimum wage on China's labor employment through empirical analysis.

(1)By collecting and analyzing the minimum wage data of 23 provinces, 5 autonomous regions and 4 municipalities directly under the Central Government, this paper studies the situation of China's minimum wage.

By analyzing the impact of the minimum wage system on labor employment, enterprise costs, economic development and industrial restructure.

(3)The panel data are used to establish an econometric model by collecting the minimum wage, employment volume, GDP and labor force population aged 16-64 in 30 regions of China from 2000 to 2022 as panel data for regression analysis, and then investigating the impact of the minimum wage on labor employment in terms of three dimensions, namely, from the national, regional and industrial perspectives.

#### 1.2.2 Research methods

This study will adopt the following three analytical methods and use the following data and tools for research:

(1) Data analysis. By collecting China's minimum wage data, this paper investigates the establishment of the two minimum wage standards by calculating and analyzing the minimum wage standard data and comparing the minimum wage standard with the monthly average wage of workers, per capita consumption expenditure, real GDP, real minimum wage standard and minimum living security standard.

(2) Empirical research and analysis. The fixed effect model is used for regression analysis to examine the overall impact of the minimum wage standard on China's labor employment from the empirical level, as well as the impact on labor employment in different regions and industries. The collected data were computationally tested through econometric methods to ensure the credibility of the findings.

(3) Empirical research and analysis. Using panel data and Hausman test,

we decide touse the fixed effect model for regression analysis to examine the overall impact o f the minimum wage standard on China's labor employment and the impact of different regions and industries on labor employment from the empirical level. Through the panel unit root test, the panel description row analysis performs a computational test on the collected data to ensure the credibility of the findings.

## 2. Basic concept of minimum wage

Minimum wage refers to the minimum labor remuneration that the employer should pay according to the law on the premise that the workers have provided normal labor during the legal working hours or the working hours agreed in the labor contract signed according to the law. Early classical economist Quesnay believed that workers should be able to earn the minimum wage necessary to maintain their livelihood, and Karl Marx also expressed some of his own views on the minimum wage. According to him, the minimum wage is the minimum amount of money that workers need to produce the goods they need to support themselves and, to some extent, their descendants. The cost of production of simple labor is the cost of maintaining the worker's existence and the continuity of the worker's offspring, and the price of this cost of maintenance and continuity is the wage, and the wage so determined is called the "minimum wage".

Minimum wage refers to the minimum labor remuneration that the employer should pay according to the law on the premise that the workers have provided normal labor during the legal working hours or the working hours agreed in the labor contract signed according to the law. The setting of China's minimum wage is a complex process, involving the comprehensive consideration of many factors.

(1) Legal basis

The minimum wage system is mainly set in accordance with the Labor Law of the People's Republic of China and the Provisions on Minimum Wages. The Labor Law provides the basic framework of the minimum wage system, while the Provisions on Minimum Wages provide detailed provisions on the setting of minimum wage standards, adjustment procedures and other relevant matters.

(2) Local governments are responsible

The minimum wage standard in China is not set uniformly by the central government but is set by the people's governments of each province, autonomous region and municipality directly under the Central Government on their own according to the local economic development level, price level, social average wage and other factors. This approach to local governance allows regions to set minimum wages more in line with local realities.

(3) Key considerations

When setting the minimum wage standard, local governments usually consider the following main factors: The average wage level of local employees: the minimum wage standard usually cannot be lower than a certain proportion of the average wage of local employees. Basic living expenses of local residents: This point mainly considers the cost of maintaining the basic living of workers, including food, housing, education and other basic expenses. Local employment situation and economic development level: this includes regional economic growth rate, labor productivity, employment rate, etc., to ensure that the minimum wage standard will not have a negative impact on employment. Social insurance premium and housing accumulation fund paid by workers individually: the actual income of workers after paying social insurance and housing accumulation fund will be considered when setting.

(4) Adjustment mechanism

According to the Provisions on Minimum Wages, the minimum wage standards shall be adjusted regularly. The people's governments of provinces, autonomous regions and municipalities directly under the Central Government generally adjust the minimum wage standards at least once every two years, but if the economic situation changes significantly, the local governments can also make timely adjustments according to the actual situation.

(5) Procedures and processes

Relevant departments of local governments will conduct extensive market research to understand the current price level, residents' living costs and enterprises' ability to pay, etc. Before the minimum wage is set or adjusted, hearings are usually held to solicit opinions from trade unions, business representatives, experts and academics. The final minimum wage is approved by provincial or municipal governments and released to the public in the form of government documents.

## 3.Current situation of China's minimum wage system

China introduced its minimum wage system in 1994, but initially, due to a lack of understanding and emphasis by local governments, it failed to effectively protect workers' rights. The turning point came with the issuance of new Minimum Wage Regulations in 2004. Since then, the minimum wage system has been in place for two decades, yet challenges remain. This chapter analyzes the current state of the minimum wage system from 2000 to 2024 and examines its impact on employment. For consistency, only monthly minimum wage data is used, and since Tibet lacked a minimum wage standard before 2004, analysis focuses on the period post-2000. Using 2010 as a pivotal point, the analysis compares the development of minimum wages before and after this year.

(1) Analysis of minimum wages before 2010

| Year / Province | 2000 | 2001 | 2002 | 2003 | 2004 | 2005 | 2006 | 2007 | 2008 | 2009 | 2010 |
|---|---|---|---|---|---|---|---|---|---|---|---|
| Beijing | 406 | 424 | 450 | 465 | 520 | 563 | 610 | 685 | 730 | 800 | 960 |
| Tianjin | 350 | 402 | 431 | 460 | 505 | 560 | 650 | 723 | 730 | 820 | 920 |
| Hebei | 290 | 290 | 340 | 350 | 435 | 520 | 535 | 580 | 750 | 750 | 900 |
| Shanxi | 300 | 300 | 340 | 340 | 430 | 520 | 528 | 565 | 645 | 645 | 850 |
| Neimenggu | 270 | 270 | 285 | 330 | 375 | 420 | 455 | 560 | 590 | 590 | 900 |
| Liaoning | 325 | 360 | 360 | 360 | 375 | 450 | 508 | 599 | 700 | 700 | 900 |
| Jilin | 425 | 270 | 297 | 335 | 360 | 360 | 460 | 580 | 600 | 600 | 820 |
| Heilongjiang | 355 | 325 | 325 | 390 | 390 | 390 | 620 | 620 | 680 | 680 | 880 |
| Shanghai | 380 | 468 | 513 | 553 | 603 | 663 | 710 | 780 | 840 | 960 | 1120 |
| Jiangsu | 275 | 410 | 445 | 500 | 690 | 690 | 705 | 775 | 850 | 850 | 960 |
| Zhejiang | 245 | 410 | 440 | 480 | 545 | 624 | 697 | 783 | 820 | 820 | 820 |
| Anhui | 320 | 340 | 355 | 370 | 380 | 410 | 438 | 530 | 560 | 560 | 720 |
| Fujian | 283 | 345 | 345 | 371 | 371 | 450 | 540 | 630 | 630 | 630 | 900 |
| Jiangxi | 245 | 250 | 250 | 250 | 287 | 360 | 366 | 510 | 580 | 580 | 720 |
| Shandong | 320 | 345 | 380 | 410 | 410 | 530 | 550 | 610 | 627 | 627 | 920 |
| Henan | 290 | 290 | 290 | 313 | 380 | 405 | 480 | 523 | 550 | 550 | 800 |
| Hunan | 220 | 225 | 245 | 300 | 350 | 412 | 412 | 508 | 635 | 635 | 850 |
| Guangdong | 450 | 460 | 485 | 510 | 525 | 684 | 716 | 780 | 780 | 1000 | 1030 |
| Shenzhen | 483 | 507 | 528 | 530 | 545 | 635 | 755 | 800 | 950 | 950 | 1100 |
| Guangxi | 200 | 257 | 340 | 340 | 360 | 460 | 473 | 513 | 580 | 670 | 820 |
| Hainan | 300 | 300 | 367 | 367 | 417 | 417 | 497 | 547 | 630 | 630 | 830 |
| Chongqing | 290 | 290 | 310 | 320 | 373 | 500 | 527 | 580 | 680 | 680 | 870 |
| Sichuan | 245 | 270 | 305 | 340 | 423 | 450 | 493 | 580 | 650 | 650 | 650 |
| Guizhou | 260 | 260 | 290 | 350 | 363 | 400 | 438 | 567 | 600 | 600 | 600 |
| Yunnan | 300 | 300 | 330 | 360 | 388 | 470 | 505 | 540 | 680 | 680 | 830 |
| Shaanxi | 260 | 275 | 320 | 320 | 320 | 405 | 503 | 540 | 540 | 540 | 760 |
| Gansu | 263 | 280 | 280 | 280 | 340 | 340 | 363 | 430 | 560 | 560 | 760 |
| Qinghai | 260 | 260 | 260 | 260 | 288 | 370 | 415 | 460 | 460 | 590 | 770 |
| Ningxia | 300 | 304 | 350 | 350 | 378 | 380 | 438 | 478 | 560 | 560 | 710 |
| Xinjiang | 390 | 408 | 460 | 460 | 473 | 480 | 547 | 580 | 670 | 670 | 800 |
| Xizang | none | | | | 470 | 470 | 470 | 470 | 680 | 680 | 730 |

Table 1 Minimum wage standards by region in China from 2000 to 2010
Source: Official website of the Ministry of Human Resources and Social Security of China

As can be seen from Table 1, the minimum wage in all Chinese provinces has been increasing year by year. Government reinforced minimum wage regulations since 2004, in the four years prior to the promulgation of the Minimum Wage Regulations, China's minimum wage increased slowly. For example, in the first-tier city of Beijing, the minimum wage increased from 406 yuan to 465 yuan between 2000 and 2003, an increase of only 59 yuan in the four years, or an average annual increase of about 4.6 percent; in the remote area of Qinghai, the minimum wage of 260 yuan had no increase in the four years; and there was no minimum wage established in Tibet. Tibet has not even established a minimum wage. In 2004, the year the Minimum Wage Regulations were promulgated, the minimum wages in most provinces were raised considerably. In Jiangsu, the minimum wage rose from 500 yuan in 2003 to 680 yuan, an increase of 180 yuan, or 12.36 percent, the largest increase; in Beijing, it rose from 465 yuan to 520 yuan, an increase of 55 yuan, or 11.8 percent, in just one year; in Shanghai, the minimum wage was raised from 553 yuan to 600 yuan, an increase of 47 yuan, or 8.5 percent; and Tibet, too, set a minimum wage of 479 yuan in that year. Tibet also set the minimum wage at 479 yuan per month that year. In the six years following the enactment of the Minimum Wage Regulations, minimum wage standards in the provinces have increased rapidly, for example, in three coastal cities, Shanghai, Guangdong and Shenzhen, where the minimum wage has risen by more than 1000 yuan; Beijing has also seen rapid growth, with an average annual increase of more than 10 percent since 2004, and Shanghai has had a large increase, with an average annual increase of about 10.3 percent. In 2010, it had reached 1120 yuan per month, the highest among all provinces.

In general, since the promulgation of the Minimum Wage Regulations in 2004, minimum wage rates in more economically developed provinces such as Shanghai, Beijing and Guangdong have increased more rapidly, and by 2010 the minimum wage rates in these areas had reached or were close to 1000 yuan. Some provinces with relatively less developed economies, such as Gansu and Ningxia, have seen relatively slower increases in minimum wage rates, but the overall trend is also upward. The increase in the minimum wage has helped to raise the income level of laborers and enhance their spending power, thereby boosting domestic demand in the economy, but it has also had an impact on the cost of labor for enterprises and on the job market

(1) Analysis of China's minimum wage from 2011 to 2024

| Year / Province | 2011 | 2012 | 2013 | 2014 | 2015 | 2016 | 2017 | 2018 | 2019 | 2020 | 2021 | 2022 | 2023 | 2024 |
|---|---|---|---|---|---|---|---|---|---|---|---|---|---|---|
| Beijing | 1160 | 1260 | 1400 | 1560 | 1720 | 1890 | 2000 | 2120 | 2200 | 2200 | 2320 | 2320 | 2320 | 2420 |
| Tianjin | 1160 | 1310 | 1500 | 1680 | 1850 | 1950 | 2050 | 2050 | 2050 | 2050 | 2180 | 2180 | 2180 | 2320 |
| Hebei | 1100 | 1320 | 1320 | 1320 | 1320 | 1650 | 1650 | 1650 | 1900 | 2020 | 1900 | 1900 | 2200 | 2200 |
| Shanxi | 980 | 1125 | 1125 | 1450 | 1620 | 1620 | 1700 | 1700 | 1700 | 1700 | 1700 | 1880 | 1980 | 1980 |
| Neimenggu | 1050 | 1200 | 1350 | 1500 | 1640 | 1640 | 1760 | 1760 | 1760 | 1760 | 1760 | 1980 | 1980 | 1980 |
| Liaoning | 1100 | 1100 | 1300 | 1300 | 1530 | 1620 | 1620 | 1810 | 1810 | 1810 | 1810 | 1910 | 1910 | 1910 |
| Jilin | 1000 | 1150 | 1150 | 1150 | 1480 | 1480 | 1780 | 1780 | 1780 | 1780 | 1780 | 1880 | 1880 | 1880 |
| Heilongjiang | 880 | 1160 | 1160 | 1160 | 1480 | 1480 | 1680 | 1680 | 1680 | 1680 | 1860 | 1860 | 1860 | 1860 |
| Shanghai | 1120 | 1450 | 1450 | 1820 | 1820 | 2190 | 2190 | 2420 | 2480 | 2480 | 2590 | 2590 | 2590 | 2690 |
| Jiangsu | 1140 | 1320 | 1480 | 1630 | 1630 | 1770 | 1890 | 2020 | 2020 | 2020 | 2280 | 2280 | 2280 | 2490 |
| Zhejiang | 1310 | 1470 | 1470 | 1650 | 1860 | 1860 | 2010 | 2010 | 2010 | 2010 | 2280 | 2280 | 2280 | 2490 |
| Anhui | 1010 | 1010 | 1260 | 1260 | 1260 | 1260 | 1260 | 1550 | 1550 | 1550 | 1550 | 1650 | 2060 | 2060 |
| Fujian | 1100 | 1200 | 1320 | 1390 | 1500 | 1600 | 1700 | 1700 | 1700 | 1800 | 1800 | 2030 | 2030 | 2030 |
| Jiangxi | 720 | 870 | 1230 | 1390 | 1530 | 1530 | 1680 | 1680 | 1680 | 1680 | 1850 | 1850 | 1850 | 2000 |
| Shandong | 920 | 1240 | 1240 | 1500 | 1500 | 1710 | 1810 | 1910 | 1910 | 1910 | 1910 | 2100 | 2100 | 2200 |
| Henan | 1080 | 1080 | 1240 | 1400 | 1400 | 1400 | 1720 | 1900 | 1900 | 1900 | 1900 | 2000 | 2000 | 2100 |
| Hunan | 1020 | 1020 | 1265 | 1265 | 1390 | 1390 | 1580 | 1580 | 1700 | 1700 | 2010 | 1930 | 2010 | 2210 |
| Guangdong | 1300 | 1300 | 1550 | 1550 | 1895 | 1895 | 1895 | 2100 | 2100 | 2100 | 1700 | 2300 | 2300 | 2300 |
| Shenzhen | 1320 | 1500 | 1600 | 1808 | 2030 | 2030 | 2130 | 2200 | 2200 | 2200 | 2100 | 2360 | 2360 | 2360 |
| Guangxi | 820 | 1000 | 1200 | 1200 | 1400 | 1400 | 1400 | 1680 | 1680 | 1810 | 2200 | 1810 | 1810 | 1990 |
| Hainan | 830 | 1050 | 1120 | 1120 | 1120 | 1430 | 1430 | 1670 | 1670 | 1670 | 1810 | 1830 | 1830 | 2010 |
| Chongqing | 870 | 1050 | 1250 | 1250 | 1250 | 1500 | 1500 | 1800 | 1800 | 1800 | 1670 | 2100 | 2100 | 2100 |
| Sichuan | 850 | 1050 | 1200 | 1400 | 1500 | 1500 | 1500 | 1780 | 1780 | 1780 | 1800 | 2100 | 2100 | 2100 |
| Guizhou | 830 | 830 | 1030 | 1250 | 1600 | 1600 | 1680 | 1680 | 1790 | 1790 | 1780 | 1790 | 1890 | 1890 |
| Yunnan | 950 | 1100 | 1265 | 1420 | 1420 | 1420 | 1420 | 1670 | 1670 | 1670 | 1790 | 1670 | 1900 | 1990 |
| Shaanxi | 860 | 1000 | 1150 | 1280 | 1480 | 1480 | 1480 | 1680 | 1800 | 1800 | 1950 | 1950 | 1950 | 2160 |
| Gansu | 760 | 980 | 1200 | 1350 | 1470 | 1470 | 1620 | 1620 | 1620 | 1620 | 1820 | 1820 | 1820 | 2020 |
| Qinghai | 920 | 1070 | 1070 | 1270 | 1270 | 1270 | 1500 | 1500 | 1500 | 1700 | 1700 | 1700 | 1880 | 1880 |
| Ningxia | 900 | 1100 | 1300 | 1300 | 1480 | 1480 | 1660 | 1660 | 1660 | 1660 | 1950 | 1950 | 1950 | 2050 |
| Xinjiang | 960 | 1340 | 1340 | 1340 | 1340 | 1340 | 1340 | 1820 | 1820 | 1820 | 1900 | 1900 | 1900 | 1900 |
| Xizang | 950 | 1200 | 1200 | 1200 | 1400 | 1400 | 1400 | 1650 | 1650 | 1650 | 1850 | 1850 | 1850 | 2100 |

Table 2 Minimum wage standards by region in China from 2010 to 2024 (average monthly wage)
Data source: the official website of the Ministry of Human Resources and Social Security of China

As can be seen from Table 2, the minimum wage in almost all provinces in China has increased to about 1000 yuan in 2011. The period from 2011 to 2015 saw the fastest increase in the minimum wage standards of each province. Among them, Gansu recorded the largest increase from 760 yuan in 2011 to 1470 yuan in 2015, representing an increase of approximately 93.4 percent; Xinjiang increased by 68.8 percent from 960 yuan to 1620 yuan in the same period; Some economically developed areas also experienced significant growth. For example, Shanghai increased from 1280 yuan to 2020 yuan, with an increase of about 57.8 percent. Beijing increased by about 48.3 percent from 1, 160 yuan to 1,720 yuan; Guangdong: from 1300 yuan to 1895 yuan, representing an increase of approximately 45.8 percent. From 2016 to 2020, the minimum wage standards of all provinces were in a period of steady growth. The increase of the minimum wage has slowed down, but it still keeps an increase of about 10 percent to 30 percent. For example, Qinghai has the largest increase of 33.9 percent from 1270 yuan in 2011 to 1700 yuan in 2020. Xizang increased by 26.9 percent from 1,300 yuan to 1,650 yuan; Beijing increased from 1,720 yuan to 2,200 yuan, an increase of about 27.9 percent. Shanghai increased from 2,020 yuan to 2,480 yuan, an increase of about 22.8 percent; Xinjiang increased by about 20.4 percent from 1,620 yuan to 1,950 yuan

Since 2011, the minimum wage system has developed rapidly and has been taken seriously by the state. The increase in the minimum wage will help workers change their quality of life, narrow the income gap between urban and rural areas, contribute to social stability, and reduce social contradictions caused by the gap between the rich and the poor.

(3) Comprehensive analysis of China's minimum wage from 2000 to 2024

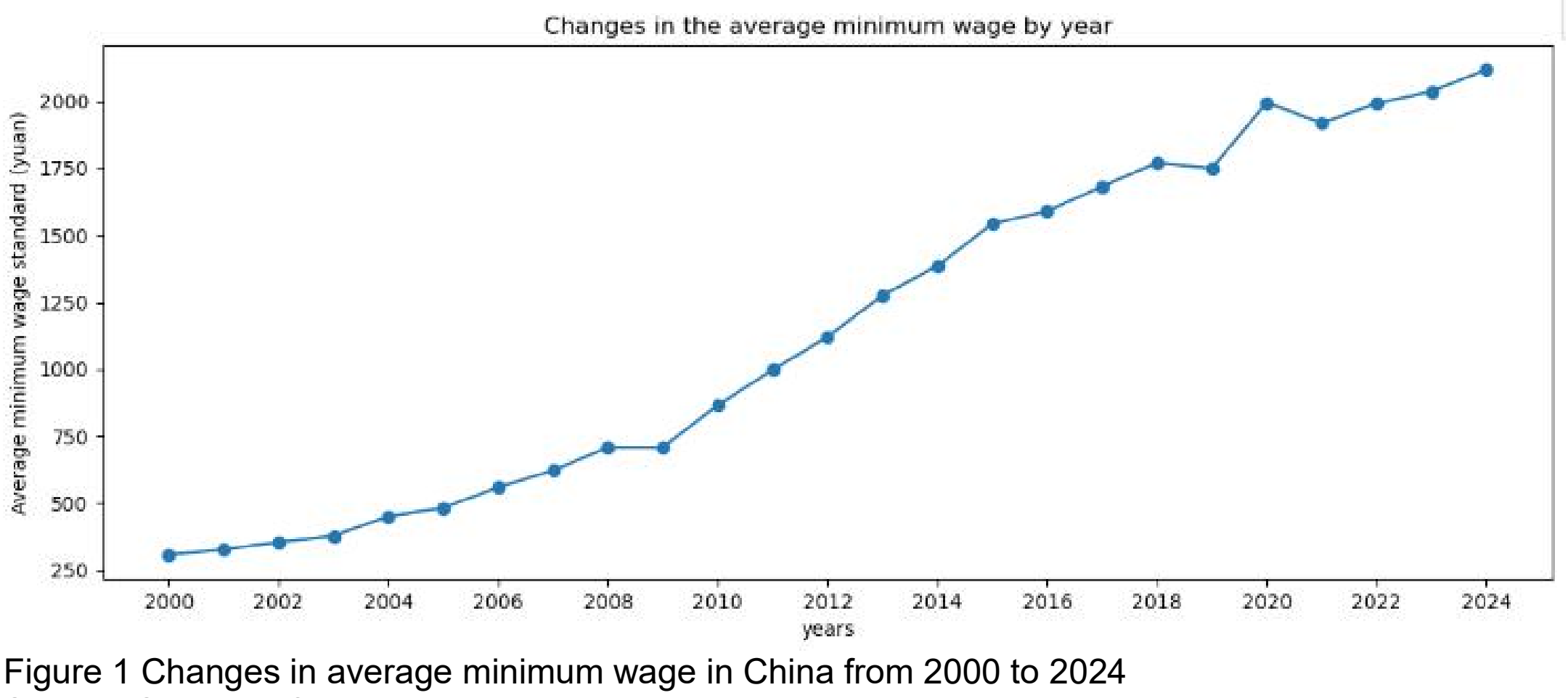

Figure 1 Changes in average minimum wage in China from 2000 to 2024
Source: Compiled from data in tables 1 and 2.

proximatAs can be seen in Figure 1, from 2000 to 2024, the minimum wage as a whole has shown a trend of continuous growth. The increase in minimum wage rates is particularly significant between 2004 and 2015. the average minimum wage was about 250 yuan in 2000, rose significantly to about 600 yuan in 2004, a large increase, continued to increase to about 1200 yuan in 2010, increased further to about 1600 yuan in 2015, reached about 1900 yuan in 2020, and is projected to reach appley 2100 yuan.

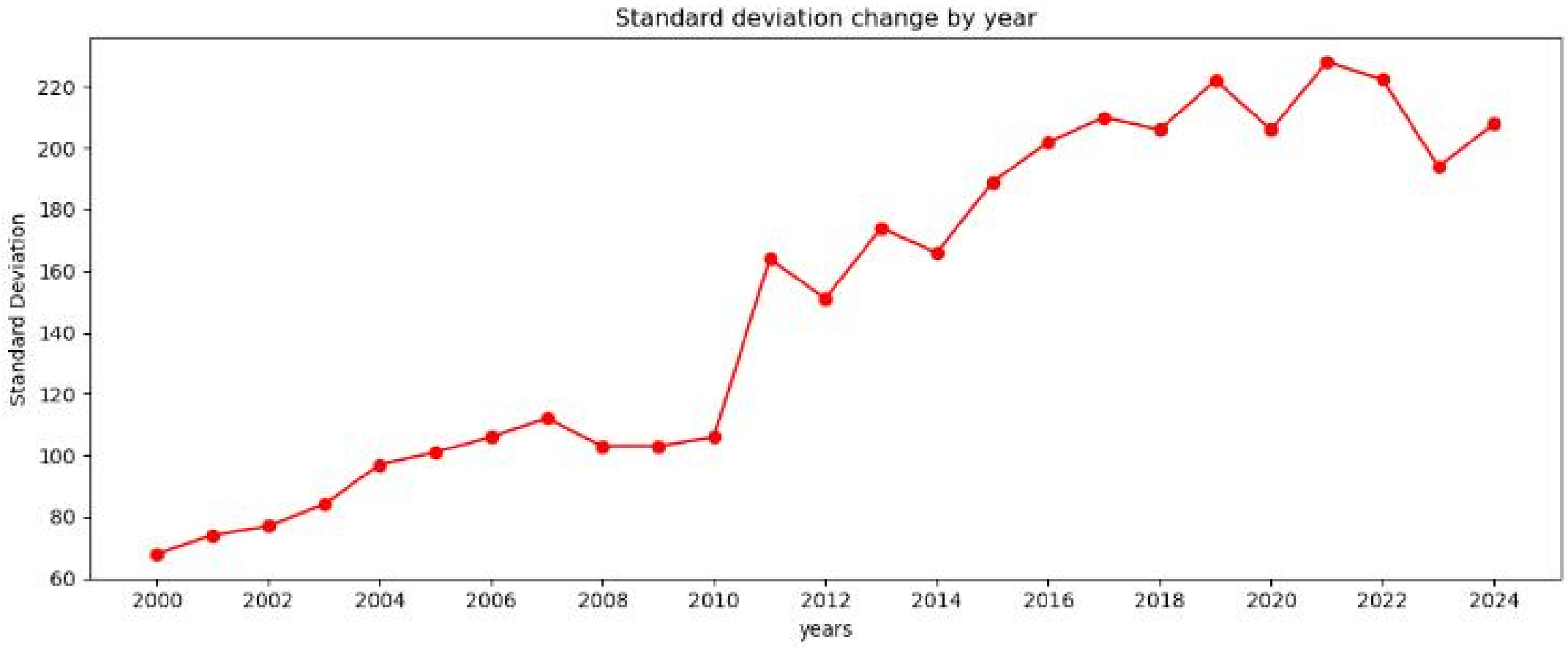

Figure 2 Changes in the standard deviation of the average minimum wage in China from 2000 to 2024

Data source: obtained from data collated from Table 1 and Table 2

The graph from Figure 2 demonstrates the changes in the standard deviation of the minimum wage standard of each province and city in China between 2000 and 2024. The standard deviation reflects the degree of dispersion of the minimum wage standard in each province and city, and the larger the standard deviation, the greater the difference of the minimum wage standard in each region.

From 2000 to 2024, the standard deviation of the minimum wage standard shows an overall upward trend. Especially after 2010, the standard deviation fluctuates more significantly, reflecting the increase in the differences in minimum wage standards across regions. from 2000 to 2010, the standard deviation gradually increased from 60 to around 100, indicating that the differences in minimum wage standards across regions began to appear, but the increase was relatively slow; from 2011 to 2012, the standard deviation increased significantly, jumping from around 100 to around 150, reflecting that the differences in minimum wage standards across regions increased during this period, and that the standard deviation of minimum wage standards across regions increased from around 100 to around 150. reflecting the increase in the difference in the adjustment rate of minimum wage standards across regions during this period; from 2013 to 2020, the standard deviation fluctuates between 150 and 200, during which the difference in minimum wage standards across regions continues to widen; and from 2021 to 2024, it is expected that the standard deviation will be around 200, reflecting the fact that the difference in minimum wage standards across regions will still remain at a relatively high level in the future.

### 3.1 Analysis of the situation of China's minimum wage standard and the average wage of employees

| Region | 2000 | | | | 2024 | | | | Average annual growth rate of the minimum wage (%) | Average monthly wage growth rate of employees (%) |
|---|---|---|---|---|---|---|---|---|---|---|
| | Monthly minimum wage | Total annual average wages | Average monthly wage | ratio | Monthly minimum wage | Total annual average wages | Average monthly wage | ratio | | |
| Beijing | 406 | 14054 | 1171 | 0.35 | 2420 | 208977 | 17415 | 0.14 | 7.87 | 12.45 |
| Tianjin | 350 | 11056 | 921 | 0.38 | 2320 | 129522 | 10794 | 0.21 | 8.28 | 11.29 |
| Hebei | 290 | 7022 | 585 | 0.50 | 2200 | 90745 | 7562 | 0.29 | 9.21 | 11.77 |
| Shanxi | 300 | 6065 | 505 | 0.59 | 1980 | 90495 | 7541 | 0.26 | 9.05 | 12.16 |
| Neimenggu | 270 | 6347 | 529 | 0.51 | 1980 | 100990 | 8416 | 0.24 | 7.57 | 12.24 |
| Liaoning | 325 | 7895 | 658 | 0.49 | 1910 | 92573 | 7714 | 0.25 | 8.25 | 11.27 |
| Jilin | 425 | 7158 | 597 | 0.71 | 1880 | 87222 | 7269 | 0.26 | 8.72 | 11.48 |
| Heilongjiang | 355 | 7094 | 591 | 0.60 | 1860 | 88235 | 7353 | 0.25 | 8.25 | 11.33 |
| Shanghai | 380 | 16641 | 1387 | 0.27 | 2690 | 212476 | 17706 | 0.15 | 8.70 | 11.65 |
| Jiangsu | 275 | 9171 | 764 | 0.36 | 2490 | 121724 | 10144 | 0.25 | 9.63 | 11.90 |
| Zhejiang | 245 | 11201 | 933 | 0.26 | 2490 | 128825 | 10735 | 0.23 | 8.91 | 11.21 |
| Anhui | 320 | 6516 | 543 | 0.59 | 2060 | 98649 | 8221 | 0.25 | 9.42 | 12.55 |
| Fujian | 283 | 9490 | 791 | 0.36 | 2030 | 103803 | 8650 | 0.23 | 9.93 | 10.96 |
| Jiangxi | 245 | 6749 | 562 | 0.44 | 2000 | 87972 | 7331 | 0.27 | 9.29 | 13.31 |
| Shandong | 320 | 7656 | 638 | 0.50 | 2200 | 102247 | 8521 | 0.26 | 8.52 | 11.93 |
| Henan | 290 | 6194 | 516 | 0.56 | 2100 | 77627 | 6469 | 0.32 | 8.76 | 11.62 |
| Hunan | 220 | 7269 | 606 | 0.36 | 2210 | 91413 | 7618 | 0.29 | 10.10 | 11.64 |
| Guangdong | 450 | 12245 | 1020 | 0.44 | 2300 | 124916 | 10410 | 0.22 | 7.35 | 10.63 |
| Shenzhen | 483 | 23039 | 1920 | 0.25 | 2360 | 171854 | 14321 | 0.16 | 6.50 | 11.18 |
| Guangxi | 200 | 6776 | 565 | 0.35 | 1990 | 92066 | 7672 | 0.26 | 10.05 | 12.01 |
| Hainan | 300 | 6865 | 572 | 0.52 | 2010 | 104802 | 8734 | 0.23 | 8.18 | 12.58 |
| Chongqing | 290 | 7182 | 599 | 0.48 | 2100 | 107008 | 8917 | 0.24 | 8.83 | 12.47 |
| Sichuan | 245 | 7249 | 604 | 0.41 | 2100 | 101800 | 8483 | 0.25 | 9.69 | 12.34 |
| Guizhou | 260 | 6595 | 550 | 0.47 | 1890 | 95410 | 7951 | 0.24 | 9.01 | 12.31 |
| Yunnan | 300 | 8276 | 690 | 0.43 | 1990 | 103128 | 8594 | 0.23 | 9.41 | 11.59 |
| Shaanxi | 260 | 6931 | 578 | 0.45 | 2160 | 98843 | 8237 | 0.26 | 9.16 | 11.18 |
| Gansu | 263 | 7427 | 619 | 0.42 | 2020 | 90870 | 7573 | 0.27 | 9.62 | 12.34 |
| Qinghai | 260 | 9081 | 757 | 0.34 | 1880 | 115949 | 9662 | 0.19 | 9.56 | 16.07 |
| Ningxia | 300 | 7392 | 616 | 0.49 | 2050 | 114631 | 9553 | 0.21 | 8.64 | 12.85 |
| Xinjiang | 390 | 7611 | 634 | 0.61 | 1900 | 101764 | 8480 | 0.22 | 7.13 | 11.94 |
| Xizang | none | 12962 | 1080 | none | 2100 | 154929 | 12911 | 0.16 | none | 11.39 |

Table 3 minimum wage and average monthly wage by region in China from 2010 to 2024
Data source: official website of the National Bureau of Statistics of the People's Republic of China

As illustrated by the aforementioned data, Hebei leads in terms of the average annual growth rate of minimum wages at 9.21 percent, whereas Xinjiang has the lowest rate at just 7.13 percent. For the average monthly wage of employees, Inner Mongolia boasts the highest growth rate at 12.78 percent, and Tianjin records the lowest with a rate of 11.29 percent.

The most substantial gap between the minimum wage and the average monthly wage of employees is observed in Xinjiang, standing at 4.81 percent. This suggests that the actual income growth for employees in Xinjiang significantly outpaced the increase in the minimum wage. Conversely, Zhejiang exhibits the smallest disparity of 1.02 percent, indicating a closer alignment between the growth of the minimum wage and the average monthly wage of employees.

Overall, the average annual growth rate of the minimum wage across most regions falls within the range of 7 to 10 percent, highlighting a substantial increase in minimum wages over the past 23 years. Meanwhile, the growth rate of average monthly wages for employees ranges from 11 to 13 percent, reflecting a notable rise in real income that surpasses the growth rate of the minimum wage.

In economically advanced areas such as Beijing and Shenzhen, the growth rate of average monthly wages for employees exceeds 12 percent, demonstrating more pronounced income growth in these regions.

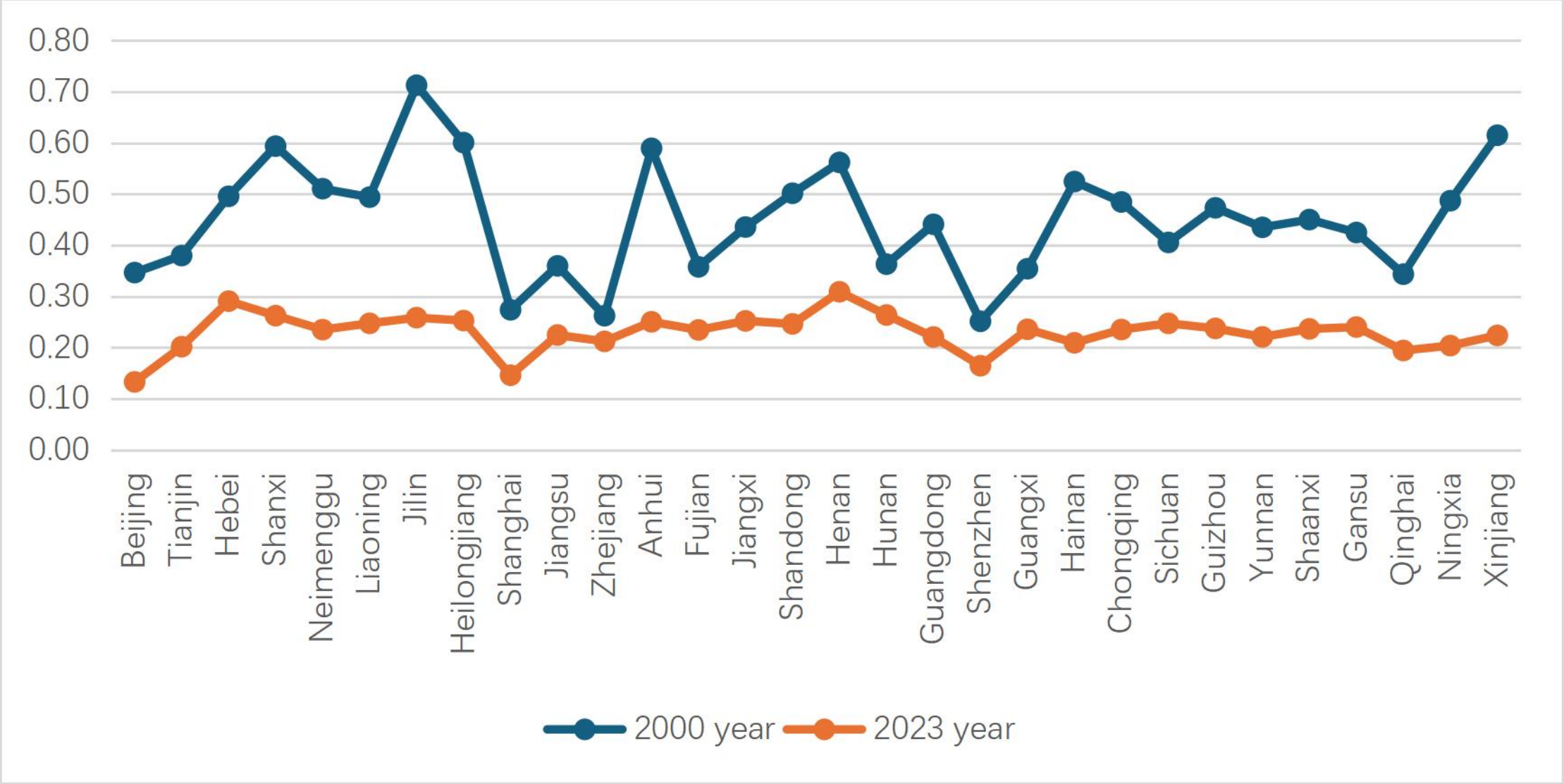

Figure 3 Ratio of the monthly minimum wage to the average monthly wage of employees by region in China in 2000 and 2023

Source: Official website of the National Bureau of Statistics of the People's Republic of China.

In 2000, the difference between the monthly minimum wage and the average monthly wage of employees in Jilin, Heilongjiang, Inner Mongolia and other regions was high, at about 0.6, indicating that the minimum wage standard in these regions was relatively high, close to half or more of the average monthly wage of employees. The ratio is lower in developed regions such as Shanghai and Beijing, especially in Shanghai, where it is only 0.27, indicating that the average monthly wages of employees in these regions are much higher than the minimum wage. By 2023, there is a significant drop in the ratio in most regions, for example, from 0.35 to 0.13 in Beijing. This suggests that despite the increase in the minimum wage, the average monthly wage of employees is increasing at a faster rate. In some regions, such as Fujian and Jiangxi, despite the decline in the ratio, the magnitude of change was relatively small, remaining between 0.2 and 0.25. In most regions, the ratio in 2023 is significantly lower than that in 2000, implying that the average wage of employees grows more rapidly as the economy develops, widening the gap with the minimum wage. The difference between the ratios of developed and less developed regions has narrowed. For example, the ratios for Shanghai and Beijing remain lower in 2023, reflecting higher average wages for employees in these regions. However, the gap with other regions is no longer as pronounced as it was in 2000. Less developed regions such as Jilin and Heilongjiang, which have higher ratios, also show a clear downward trend. In some central regions, such as Henan and Hubei, the ratio was relatively stable at both points in time, reflecting the relatively stable economic development and wage growth in these regions.

### 3.2 Analysis of China's minimum wage standard and the per capita consumption level of urban residents

| Region | 2000 | | | 2023 | | |
|---|---|---|---|---|---|---|
| | Monthly minimum wage | The average monthly consumption expenditure of urban households per person | difference value | Monthly minimum wage | The average monthly consumption expenditure of urban households per person | difference value |
| Beijing | 406 | 708 | 302 | 2320 | 3557 | 1237 |
| Tianjin | 350 | 510 | 160 | 2180 | 2610 | 430 |
| Hebei | 290 | 362 | 72 | 1900 | 1741 | -159 |
| Shanxi | 300 | 328 | 28 | 1880 | 1461 | -419 |
| Neimenggu | 270 | 327 | 57 | 1980 | 1858 | -122 |
| Liaoning | 325 | 363 | 38 | 1910 | 1884 | -26 |
| Jilin | 425 | 335 | -90 | 1880 | 1491 | -389 |
| Heilongjiang | 355 | 319 | -36 | 1860 | 1701 | -159 |
| Shanghai | 380 | 739 | 359 | 2590 | 3837 | 1247 |
| Jiangsu | 275 | 444 | 169 | 2280 | 2737 | 457 |
| Zhejiang | 245 | 585 | 340 | 2280 | 3248 | 968 |
| Anhui | 320 | 353 | 33 | 1650 | 1878 | 228 |
| Fujian | 283 | 470 | 187 | 2030 | 2503 | 473 |
| Jiangxi | 245 | 302 | 57 | 1850 | 1809 | -41 |
| Shandong | 320 | 419 | 99 | 2100 | 1887 | -213 |
| Henan | 290 | 319 | 29 | 2000 | 1585 | -415 |
| Hunan | 220 | 435 | 215 | 1930 | 2069 | 139 |
| Guangdong | 450 | 668 | 218 | 2300 | 2681 | 381 |
| Shenzhen | 483 | 446 | -37 | 2360 | 2451 | 91 |
| Guangxi | 200 | 404 | 204 | 1810 | 1529 | -281 |
| Hainan | 300 | 340 | 40 | 1830 | 1792 | -38 |
| Chongqing | 290 | 464 | 174 | 2100 | 2114 | 14 |
| Sichuan | 245 | 405 | 160 | 2100 | 1858 | -242 |
| Guizhou | 260 | 357 | 97 | 1790 | 1495 | -295 |
| Yunnan | 300 | 432 | 132 | 1670 | 1579 | -91 |
| Shaanxi | 260 | 356 | 96 | 1950 | 1654 | -296 |
| Gansu | 263 | 344 | 81 | 1820 | 1457 | -363 |
| Qinghai | 260 | 349 | 89 | 1700 | 1438 | -262 |
| Ningxia | 300 | 350 | 50 | 1950 | 1595 | -355 |
| Xinjiang | 390 | 369 | -21 | 1900 | 1494 | -406 |
| Xizang | none | 463 | none | 1850 | 1324 | -526 |

Table 4 The difference between the monthly minimum wage and the average monthly consumption expenditure of urban households in 2000 and 2023

In the year 2000, regions such as Jilin, Heilongjiang, and Inner Mongolia exhibited a high difference between the monthly minimum wage and the average monthly wage of employees, approximately 0.6. This indicated that the minimum wage standards in these areas were relatively elevated, often reaching close to half or more of the average monthly wage. Conversely, developed regions like Shanghai and Beijing showed lower ratios, particularly Shanghai with a ratio of only 0.27, highlighting much higher average monthly wages for employees compared to the minimum wage.

By 2023, most regions experienced a significant decline in this ratio. For instance, Beijing's ratio dropped from 0.35 to 0.13, suggesting that although the minimum wage increased, the average monthly wage of employees grew at a faster pace. In some provinces, such as Fujian and Jiangxi, despite a decrease in the ratio, the change was relatively modest, staying within the range of 0.2 to 0.25.

Overall, the ratio in 2023 is notably lower than in 2000 for most regions, indicating that the average wage of employees has grown more rapidly as economies have developed, thereby widening the gap with the minimum wage. The disparity between developed and less developed regions has narrowed. While ratios in Shanghai and Beijing remain low in 2023, reflecting higher average employee wages in these regions, the gap with other regions is not as pronounced as it was two decades ago.

Less developed regions, such as Jilin and Heilongjiang, which initially had higher ratios, also demonstrated a clear downward trend. Meanwhile, central regions like Henan and Hubei maintained relatively stable ratios over the two time points, indicative of steady economic development and wage growth in these areas.

This analysis underscores the dynamic changes in wage structures across different regions of China, reflecting broader trends in economic development and income distribution.

### 3.3 Analysis of Minimum Wage and GDP per capita in China

Gross Domestic Product (GDP) is an important economic indicator who is able to measure the economic activity and economic health of a country or region over a specific period of time. Thus, it provides valuable information for various aspects such as policy making, employment, investment, international trade and quality of life. Gross Domestic Product per capita (GDP per capita) is an indicator obtained by dividing the total GDP of a country or region by the total population of that country or region. It is commonly used to measure the average economic productivity and standard of living of a country or region. GDP per capita provides a better perspective on the quality of life of the population because it takes into account demographic factors.

| Region | GDP per capita in 2000 (Yuan) | GDP per capita in 2023 (Yuan) | Average annual growth rate of the minimum wage (%) | Average annual growth rate of GDP per capita (%) |
|---|---|---|---|---|
| Beijing | 22460 | 190313 | 7.87 | 9.74 |
| Tianjin | 17993 | 119235 | 8.28 | 8.57 |
| Hebei | 7663 | 56995 | 9.21 | 9.12 |
| Shanxi | 5137 | 73675 | 9.05 | 12.28 |
| Neimenggu | 5872 | 96474 | 7.57 | 12.94 |
| Liaoning | 11226 | 68775 | 8.25 | 8.20 |
| Jilin | 6847 | 55347 | 8.72 | 9.51 |
| Heilongjiang | 8562 | 51096 | 8.25 | 8.08 |
| Shanghai | 34547 | 179907 | 8.70 | 7.44 |
| Jiangsu | 11773 | 144390 | 9.63 | 11.51 |
| Zhejiang | 13461 | 118496 | 8.91 | 9.92 |
| Anhui | 4867 | 73603 | 9.42 | 12.54 |
| Fujian | 11601 | 126829 | 9.93 | 10.96 |
| Jiangxi | 4851 | 70923 | 9.29 | 12.37 |
| Shandong | 9555 | 86003 | 8.52 | 10.02 |
| Henan | 5444 | 62106 | 8.76 | 11.16 |
| Hunan | 5639 | 73598 | 10.10 | 11.82 |
| Guangdong | 12885 | 101905 | 7.35 | 9.41 |
| Shenzhen | 32800 | 183172 | 6.50 | 7.76 |
| Guangxi | 4319 | 52164 | 10.05 | 11.44 |
| Hainan | 6894 | 66602 | 8.18 | 10.36 |
| Chongqing | 5157 | 90663 | 8.83 | 13.27 |
| Sichuan | 4784 | 67777 | 9.69 | 12.22 |
| Guizhou | 2662 | 52321 | 9.01 | 13.83 |
| Yunnan | 4637 | 61716 | 9.41 | 11.91 |
| Shaanxi | 4549 | 82864 | 9.16 | 13.45 |
| Gansu | 3838 | 44968 | 9.62 | 11.29 |
| Qinghai | 5087 | 60724 | 9.56 | 11.38 |
| Ningxia | 4839 | 69781 | 8.64 | 12.30 |
| Xinjiang | 7470 | 68552 | 7.13 | 10.12 |
| Xizang | 4559 | 58438 | none | 11.73 |

Table 5 Annual average growth rates of minimum wage and GDP per capita in China
Data source: Official website of the National Bureau of Statistics of the People's Republic of China

According to the table, the highest average annual growth rate of per capita GDP is observed in Guizhou at 13.83 percent, followed closely by Shaanxi at 13.45 percent and Chongqing at 13.27 percent. In contrast, Shenzhen and Shanghai exhibit relatively lower growth rates of 7.76 percent and 7.44 percent, respectively. Despite their robust economic foundations, these two regions show a more moderate pace in the increase of per capita GDP.

The average annual growth rate of the minimum wage standard aligns closely with the per capita GDP growth rate in provinces like Zhejiang, Anhui, and Fujian. However, in Shanxi and Inner Mongolia, the minimum wage standard's growth rate lags significantly behind the per capita GDP growth, indicating that the minimum wage has increased more slowly relative to the rapid economic expansion in these regions.

From a regional economic development perspective, the developed eastern coastal areas, including Beijing, Shanghai, Jiangsu, and Zhejiang, boast higher levels of per capita GDP but experience relatively stable and moderate growth rates. Conversely, less developed regions in central and western China, such as Guizhou, Shaanxi, and Chongqing, start from a lower GDP per capita base but achieve higher growth rates. This pattern reflects the dynamic economic transformation occurring across different parts of the country, with less developed regions catching up at a faster pace.

Overall, the average annual growth rate of per capita GDP (represented by the orange line) generally exceeds that of the minimum wage (depicted by the blue line). Most regions experience an average annual minimum wage growth rate between 7-10 percent, while the per capita GDP growth rate typically ranges from 8-14 percent. This trend suggests that during the period from 2000 to 2003, China's economy expanded at a relatively rapid pace, with wage growth lagging behind economic expansion.

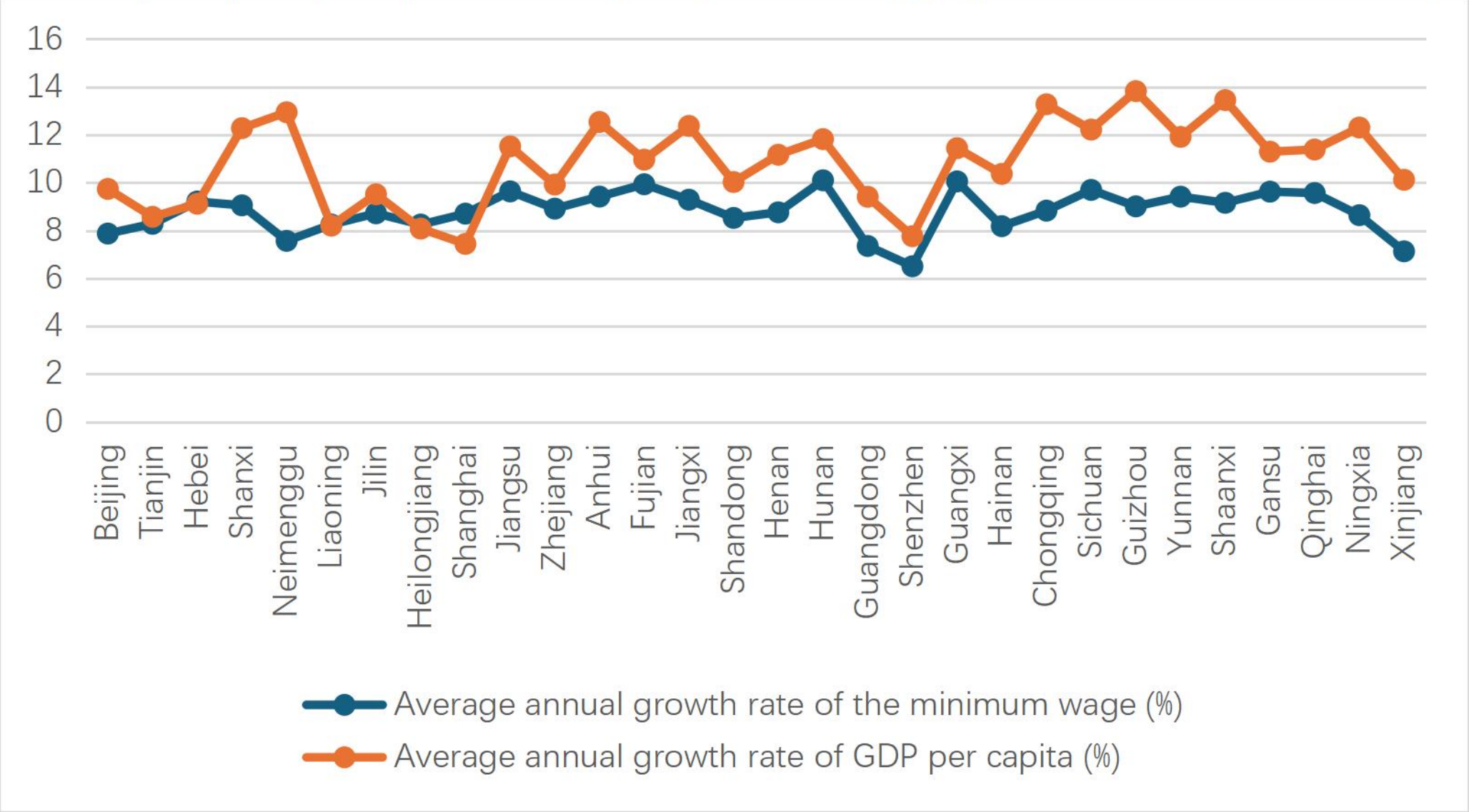

Figure 4 Annual average growth rates of minimum wage and GDP per capita in China
Data source: collated from Table 5

In particular, provinces such as Shanxi, Inner Mongolia, Chongqing, Guizhou, and Shaanxi exhibit significantly higher per capita GDP growth rates compared to their minimum wage standards. This indicates that in these regions, rapid economic growth outpaced wage increases. Conversely, in provinces like Hunan, Guangxi, and Yunnan, the growth rates of the minimum wage standard and per capita GDP are more closely aligned, suggesting that economic growth is more effectively translating into wage gains for workers.

On the other hand, first-tier cities including Beijing, Tianjin, Shanghai, Guangdong, and Shenzhen show relatively lower growth rates for both per capita GDP and minimum wages. Especially in Shenzhen and Shanghai, despite their strong economic foundations, growth has tended to be more moderate. This pattern reflects the limited growth potential characteristic of economically mature stages in these regions.

### 3.1.4 Analysis of China's minimum wage standard and monthly minimum living standard for urban residents

The monthly minimum subsistence guarantee standard for urban residents is a social assistance system that refers to the minimum subsistence guarantee payment standard set by the government to ensure the basic livelihood of residents who are unable to maintain a basic life through their own income or low-income residents, with the aim of ensuring that the poorest urban residents are able to satisfy their basic living needs, including food, housing, and medical care. By means of this direct subsidy, the Government helps these residents to escape from extreme poverty, reduces social inequality, promotes social equity and reduces social problems caused by poverty, such as rising crime rates and social unrest. Although the setting of a minimum standard of living guarantees the basic needs of these people, it is also necessary to avoid the phenomenon of "supporting lazybones" as a result of an excessively high level of protection. For this reason, the Government will usually set up relevant measures to encourage the protected people to take up active employment. The minimum wage and the low-income protection system complement each other. The minimum wage guarantees a minimum level of income for workers, while the minimum living standard provides protection for those who cannot earn enough income through labor. By comparing the monthly minimum subsistence guarantee standard for urban residents with the minimum wage, it is possible to gain a fuller understanding of the Government's dual regulatory role in social security and the labor market, and thus to formulate more rational and effective socio-economic policies.

As can be seen from the above table, the LPS standard in most regions is significantly lower than the minimum wage standard. For example, Beijing's LPS standard is 1395 yuan, while the minimum wage is 2320 yuan; Ningxia's LPS standard is the lowest at 460 yuan, while the minimum wage is 1950 yuan. This disparity ensures that workers are incentivized to work by receiving higher incomes when they are employed. Economically developed regions such as Beijing, Shanghai, and Guangdong have higher LPS and minimum wage standards, reflecting higher costs of living and economic levels. While less economically developed regions such as Ningxia and Qinghai have lower standards.

| Region | Monthly minimum wage standard（yuan） | Monthly minimum living allowances for urban residents（yuan） |
| --- | --- | --- |
| Beijing | 2320 | 1395 |
| Tianjin | 2180 | 780 |
| Hebei | 2200 | 850 |
| Shanxi | 1980 | 760 |
| Neimenggu | 1980 | 840 |
| Liaoning | 1910 | 747 |
| Jilin | 1880 | 715 |
| Heilongjiang | 1860 | 689 |
| Shanghai | 2590 | 710 |
| Jiangsu | 2280 | 800 |
| Zhejiang | 2280 | 1100 |
| Anhui | 2060 | 788 |
| Fujian | 2030 | 900 |
| Jiangxi | 1850 | 885 |
| Shandong | 2100 | 944 |
| Henan | 2000 | 630 |
| Hunan | 2010 | 650 |
| Guangdong | 2300 | 1238 |
| Shenzhen | 2360 | 1413 |
| Guangxi | 1810 | 810 |
| Hainan | 1830 | 690 |
| Chongqing | 2100 | 735 |
| Sichuan | 2100 | 740 |
| Guizhou | 1890 | 737.5 |
| Yunnan | 1900 | 728 |
| Shaanxi | 1950 | 780 |
| Gansu | 1820 | 920 |
| Qinghai | 1880 | 703 |
| Ningxia | 1950 | 460 |
| Xinjiang | 1900 | 678 |
| Xizang | 1850 | 947 |

Table 6 China's Minimum Wage Standards and Monthly Minimum Subsistence Guarantee Standards for Urban Residents
Source: Official websites of Chinese provincial and municipal governments

## 4. Empirical Analysis of the Impact of Minimum Wage on Labor Force Employment

This chapter will further study the impact of the minimum wage on employment from the empirical level and explain the empirical results by constructing an econometric model.

### 4.1 National Analysis

#### 4.1.1 Model setting

Empirical calculation of the impact of minimum wage on employment also frequently employs a simple equation of regression in which employment is the dependent variable and minimum wage is the most significant independent variable. Other control variables such as GDP and labor supply (LS) are also used in this equation to allow for adjustment in view of other factors influencing employment. The overall form of the regression equation is as follows:

$$E = \alpha(MW) + \beta_1 GDP + \beta_2 LS + \epsilon$$

Where E (employment ) represents the employment volume, MW (minimum wage) represents the minimum wage standard, GDP represents the growth rate of GDP, LS represents the working-age population aged 16-64, and ϵ is the error term.

This model structure has been commonly applied in empirical work on minimum wage impacts on employment in many countries. It examines the impact of changes in minimum wages on jobs while controlling for macroeconomic conditions and demographic factors. Empirical evidence used here also extends models of Neumark (2001) and Dickens, Machin, and Manning (1999), who applied similar regression techniques to investigate the minimum wage-jobs relationship in different regions. These studies contributed a lot to clarifying the economic impacts of minimum wage policies, with the necessity of considering economic growth and labor market conditions being emphasized.

The model is modified in this paper to fit the unique conditions in China by considering regional-specific economic conditions as well as labor force characteristics. The new model is specified as follows:

$$\Delta lnE_{it} = \alpha + \beta_1 \Delta lnMW_{it} + \beta_2 \Delta lnX_{it} + \epsilon_{it}$$

Where $E_{it}$ represents the logarithmic difference of the employment change of region i in period t; ΔlnMw it represents the logarithmic difference of the change of the minimum wage standard in region i in period t; Δlnx it is the log difference of other control variables (such as regional GDP); α is the constant term and $E_{it}$ is the error term.

By considering the actual situation in China, the above model is adjusted and specifically set as follows:

$$\Delta lnE_{it} = \alpha + \beta_1 \Delta lnMW_{it} + \beta_2 \Delta lnGDP_{it} + \beta_3 \Delta lnLS_{it} + \mu_{it} + \epsilon_{it}$$

(1) Level of employment E it : The level of employment in every region is indicated by the number of employed individuals gathered, including those aged 16 years and older who are involved in some social labor and earn labor remuneration or business income. It can actually quantify the real utilization of all the labor forces within a specific time span and capture the effect of the minimum wage policy on the labor market directly. It should be noted that the level of employment is used as the dependent variable to analyze the impact of the minimum wage on employment rather than the amount of unemployment, mainly due to the following reasons: Firstly, the employment measure captures the full picture of the labor market by including all those who are employed, while the unemployment measure only considers those who are not working but looking for work and can possibly overlook some informal employment or labor market dropouts. Second, the employment level analysis can more confidently ensure data reliability, as employment data are generally more complete and accurate than unemployment data, especially in some countries or areas. Moreover, the employment number directly indicates the impact of the minimum wage policy on the real demand for labor, whereas the unemployment number may be subject to a number of multiple factors, thus it is not possible to compute the specific impact of the minimum wage policy alone. In addition, labor market and informal economy mechanisms are better captured by employment than unemployment: minimum wage hikes can move workers from formal to informal employment. Finally, economic growth in employment is one of the core goals of economic policy, and an examination of the employment level is more in line with the concern of the policy maker and supports policy-making and policy adjustment.

(2) The rate of minimum wage Mw it: All provinces and cities in China have set the minimum wage standard, but some have set multiple grades. Taking the top bracket, the policy impact on a larger number of workers than just the lowest earners can be assessed. Second, the level of the highest grade reflects the general level of the region or industry and can be more representative of the overall economic level and the minimum level of protection in the labor market, and hence it is more widely representative to a certain degree. In addition, the highest level of minimum wage tends to be higher than any other level, and the firm cost impact is greater. Therefore, the use of the highest level is useful to measure the "ceiling" effect of minimum wage policy, i.e., the impact of the policy on employment and enterprise cost under the most stringent conditions.

This model captures the effect of minimum wage changes on local employment growth after controlling for other macroeconomic variables. The choice of using employment quantity as the dependent variable, over unemployment, is deliberate. Employment levels provide a wider picture of labor market performance since it captures the number of workers in the overall economy, including those in the informal economy, which may be left out by unemployment measures.

In order to make the model stronger, this study examines the highest minimum wage rate by region. This approach gives a broader view of the impact of the policy since it is sure to look at the impact across a greater number of workers and provide a better picture of the regional economy.

#### 4.1.2 Data Description

Due to data limitation, empirical analysis of this study uses time-series data. While time-series data can be helpful, it may not be sufficient to draw complete inferences for different regions with different economic status. Hence, the study focuses on regions for which complete data are available. Tibet is excluded due to unavailability of minimum wage data.

Minimum wage data are collected from China Labor and Personnel Network and Labor Consulting Network. The rest of the economic and demographic data like GDP and labor supply are collected from China Statistical Yearbook and local statistical yearbooks of the National Bureau of Statistics.

4.1.3 Empirical test

1) Levin-Lin-Chu Test (LLS)

The purpose of the LLC test is to judge whether the panel data has a unit root, that is, whether there is a co-root process. The LLC test is based on the following panel data regression model:

$$\Delta y_{it} = \alpha_i + \beta y_{it-1} + \sum_{k=1}^{p} \gamma_{ik} \Delta y_{it-k} + \epsilon_{it}$$

Where $y_{it}$ is the observed value of cross-section unit i at time t. $\alpha_i$ is the cross-sectional fixed effect. β is the key parameter of the test, if β = 0, there is a unit root; If β< 0, then the series is stationary. $\gamma_{ik}$ is the lag term coefficient and $\Delta y_{it-k}$ is used to control for the autocorrelation of the series. The $\epsilon_{it}$ is the error term.

The null hypothesis, H0, is the existence of a unit root, and the alternative hypothesis, H1, is that it is stationary. We conclude that the data is stationary if the test statistic is strongly negative and the p-value is smaller than 0.05.

(2) Im-Pesaran-Shin test (IPS)

The IPS test differs from the LLC test in that the IPS test allows the unit root process to be heterogeneous across cross-section units, that is, each cross-section unit can have different unit root properties. This feature makes the IPS test more flexible when analyzing panel data with heterogeneous cross-sectional units.

IPS test is based on the following panel data regression model:

$$\Delta y_{it} = \alpha_i + \beta_i y_{it-1} + \sum_{k=1}^{p} \gamma_{ik} \Delta y_{it-k} + \epsilon_{it}$$

Where yit is the observed value of cross-section unit i at time t. $\alpha_i$ is the cross-sectional fixed effect. βi is the unit root parameter of cross section unit i. If βi = 0, it means that there is a unit root. If βi <0, then the series is stationary. $\gamma_{ik}$ is the lag term coefficient and $\Delta y_{it-k}$ is used to control for the autocorrelation of the series. The $\epsilon_{it}$ is the error term.

Null hypothesis (H0): there is a unit root in all cross-sectional units, that is, the panel data is not stationary as a whole.

Alternative hypothesis (H1): There is no unit root in at least one of the cross-sectional units, that is, at least one of the cross-sectional units is stationary.

Judgment Criteria: If the p-value is less than the significance level (generally 0.05 or 0.01), then at least one cross-sectional unit is non-stationary. If the p-value is greater than or equal to the significance level, then the entire panel is non-stationary.

3) Augmented Dickey-Fuller Test

The ADF test is an extended version of the Dickey-Fuller test, which addresses the issue of serial autocorrelation by adding lagged terms.

$$\Delta y_t = \alpha_i + \beta y_{it-1} + \lambda_t + \sum_{i=1}^{p} \gamma_i \Delta y_{t-i} + \epsilon_t$$

Where yt is the observed value at time t. Δyt = yt - yt−1 is the difference term of the series, which is used to eliminate the autocorrelation of the series. $\alpha_i$ is the constant term (the intercept). β is the unit root parameter tested. If β = 0, it means that the sequence has a unit root. If β < 0, then the series is stationary. Is the coefficient of the time trend. The λt is the coefficient of the lagged term, which is used to control for higher order autocorrelation of the series. So $\epsilon_t$ is the error term.

Null hypothesis (H0): The series has unit root (i.e., the series is not stationary)

Alternative Hypothesis (H1): There is no unit root in the series (i.e., the series

is stationary)

Judgment criteria: If the statistic of the ADF test is less than the critical value, or the p-value is less than the significance level (usually 0.05), the null hypothesis is rejected and the series is considered stationary, that is, there is no unit root. If the p-value of the ADF test is greater than the significance level, the null hypothesis cannot be rejected and the series is considered to have a unit root, that is, it is not stationary.

(2) Unit root test results

As seen in the table below, the results of the three unit root tests (Levin-Lin-Chu (LLC) test, Im-Pesaran-Shin (IPS) test, and ADF-Fisher test) indicate that the variables exhibit varying behaviors towards stationarity, which is the most critical phase in panel data analysis. LLC test is based on the homogeneous panel data assumption where all cross-sectional units share a common unit root characteristics. The IPS test, however, allows for heterogeneous unit root behavior across cross-sectional units, which provides more flexibility working with differentiated panels. Lastly, the ADF-Fisher test extends the standard Augmented Dickey-Fullard (ADF) test to allow for panel data and provides a more general framework for stationarity testing.

Levin-Lin-Chu Test (LLC):

The LLC test results reveal that the variables E, MW, GDP, and LS are all not unit rooted, as evidenced by the extremely negative test statistics and the very low p-values (much less than 0.05). This means that the variables are stationary, i.e., data does not have a trend or systematic pattern over time, and we can go ahead with further analysis assuming stationarity for these variables. Specifically, lnE, lnMW, lnGDP, and lnLS all satisfy the LLC test for stationarity with negative test statistics and close to zero p-values (high statistical significance).

Im-Pesaran-Shin Test (IPS):

The IPS test result indicates that the MW variable is non-stationary in level form, since the test statistic is positive while the p-value is huge (0.756). On the contrary, variables E, GDP, and LS are stationary, since the test statistics are negative with their corresponding p-values of below 0.05. The test, however, goes further to state that E and LS still have a unit root on the 1% level of significance, recognizing that the unit root test may be sensitive with respect to different model specifications. MW is positive for the test statistic and 0.756 for p-value, that it fails to pass the IPS stationarity test and is thus compliant with LLC. It indicates MW requires differencing or transformation in order to attain stationarity.

ADF-Fisher Test:

The ADF-Fisher test also confirms the findings from the LLC and IPS tests. It informs us that E and MW are non-stationary in levels because they have unit roots. Their p-values are huge, indicating failure to reject the null hypothesis of a unit root. On the other hand, GDP and LS are stationary because their p-values are extremely low, indicating rejection of the null hypothesis of a unit root. The ADF-Fisher test emphasizes the observation that although GDP and LS are stationary in their level structures, the data transformation of E and MW (e.g., differencing) is required to achieve stationarity.

Conclusion on Stationarity:

After the application of differencing (i.e., ΔlnE, ΔlnMW, ΔlnGDP, and ΔlnLS), it is observed that all the variables are discovered to be stationary in all the three tests. Stationarity of the variables after differencing is critical in time series analysis and regression because non-stationarity data may produce spurious results. MW, for instance, was not stationary at its level form under both the LLC and IPS tests but becomes stationary after differencing, confirming that the non-stationary variables should be differenced before proceeding with further analysis.

| Variable | LLC test | Prob | IPS test | Prob | ADF-Fisher test | Prob | Stationary/Non-stationary |
|---|---|---|---|---|---|---|---|
| lnE | -6.685 | 0.000 | -1.896 | 0.029 | 85.740 | 0.158 | Stationary |
| ΔlnE | -8.081 | 0.000 | -8.837 | 0.000 | 233.356 | 0.000 | Stationary |
| lnMW | -6.543 | 0.000 | 0.695 | 0.756 | 44.382 | 0.934 | Non-Stationary |
| ΔlnMW | -18.439 | 0.000 | -17.006 | 0.000 | 366.195 | 0.000 | Stationary |
| lnGDP | -13.406 | 0.000 | -5.578 | 0.000 | 136.504 | 0.000 | Stationary |
| ΔlnGDP | -10.726 | 0.000 | -10.563 | 0.000 | 299.630 | 0.000 | Stationary |
| lnLS | -4.290 | 0.000 | -1.713 | 0.043 | 95.899 | 0.002 | Stationary |
| ΔlnLS | -16.559 | 0.000 | -14.306 | 0.000 | 334.365 | 0.000 | Stationary |

Table 7 Unit root test for panel data

(3) Fixed effects model test results

The result of Hausman's test can be used to decide whether to apply the fixed effects model or random effects model in panel data analysis. This method helps to get rid of the cross-section heteroskedasticity that may be caused by, among other factors, different economic development levels in different regions, and the result of the Cross-section and period random test shows that the Chi-Sq statistic is 73.744503 and p-value is 0.0000, which shows that if we have both time and cross-section in consideration, there is a very strong difference between the fixed-effects model and the random-effects model. fixed effects models differ significantly from each other. Since the p-value is significantly lower than 0.05, the random effects model is rejected in favor of the fixed effects model.

| Test Summary | Chi-Sq. Statistic | Chi-Sq. d.f. | Prob. |
|---|---|---|---|
| Cross-section random | 0.00 | 3.00 | 1.00 |
| Period random | 0.00 | 3.00 | 1.00 |
| Cross-section and period random | 73.74 | 3.00 | 0.00 |

Table 8 Results of Hausman test

**4.1.4 Correlation analysis**

The correlation test in the table displays several key relationships between economic variables. Labor force population (LNLS) and employment (LNE) have a strong positive relationship as indicated by the correlation coefficient of 0.9741. This indicates that when labor force population rises, the rate of employment also rises significantly, meaning that labor force population is a significant determinant of employment. In addition, the positive correlation between GDP (LNGDP) and employment (LNE) is high with a coefficient of 0.6583. It reflects that economic growth is usually accompanied by an increase in employment levels as economic growth creates more demand for labor. Minimum wage (LNMW) and GDP (LNGDP) also have a high positive correlation, with a coefficient of 0.7465. This picks up on the tendency for the minimum wage level to rise with economic expansion, possibly catching up with policy aims to ensure that living levels at the basics for workers are preserved during periods of high economic performance. The relationship between minimum wage (LNMW) and employment (LNE) and labor force population (LNLS) is, however, quite weak, with coefficients of 0.0980 and 0.0728, respectively. This suggests that changes in the minimum wage have little direct impact on aggregate labor supply and employment.

Overall, the findings suggest that the level of employment is primarily determined by economic growth and the size of the labor force, and the minimum wage is more strongly related to economic growth but less so to labor supply and employment. Secondly, the relationship between differenced variables (e.g., ΔLNE, ΔLNMW, etc.) is mostly weak, indicating that the short-run fluctuations in GDP and employment are more indirectly related and may be influenced by other factors.

| | LNE | LNMW | LNGDP | LNLS | ΔLNE | ΔLNMW | ΔLNGDP | ΔLNLS |
|---|---|---|---|---|---|---|---|---|
| LNE | 1.0000 | | | | | | | |
| LNMW | 0.0980 | 1.0000 | | | | | | |
| LNGDP | 0.6583 | 0.7465 | 1.0000 | | | | | |
| LNLS | 0.9741 | 0.0728 | 0.6499 | 1.0000 | | | | |
| ΔLNE | -0.0650 | -0.1209 | -0.0905 | -0.0840 | 1.0000 | | | |
| ΔLNMW | 0.0306 | -0.0410 | -0.0293 | 0.0288 | 0.1289 | 1.0000 | | |
| ΔLNGDP | -0.0226 | -0.2220 | -0.1173 | -0.0339 | 0.1871 | 0.2575 | 1.0000 | |
| ΔLNLS | -0.0731 | -0.1399 | -0.0995 | -0.0683 | 0.1634 | 0.1350 | 0.1897 | 1.0000 |

Table 9 Correlation results of variables

#### 4.1.5 Regression analysis of the impact of minimum wage on employment: nationwide

The regression in Table 10 shows the effect of minimum wage on employment, along with the impact of GDP and labor supply.

The regression estimate for minimum wage (LNMW) is -0.2074 with a t-statistic of -5.8609 and p-value of 0.0000. This indicates that minimum wage has a significant negative association with employment. Accurately, 1% increase in the minimum wage has a relationship of -0.2074% with employment falling. This negative relationship can be explained on account of the fact that firms reduce recruitment as a consequence of higher labor costs brought about by a rise in the minimum wage. That is, while there is a rise in minimum wages, firms can try to minimize their workforce in an effort to remain profitable. The gross domestic product coefficient (LNGDP) is -0.0645 with t-statistic of -3.1423 and p-value of 0.0018, which represents that GDP and employment have a negative correlation at the significance level. While GDP in most cases represents economic growth, in this case the negative coefficient shows that growth in the economy may reduce employment. This can be due to structural changes in the economy, whereby GDP growth is driven by more capital-intensive industries that fail to generate jobs proportionally, or merely automation and other technological improvements that reduce the need for labor.

On the other hand, labor supply (LNLS) also depicts a high and positive correlation with employment. Its coefficient is 0.8490 with the t-statistic equal to 18.5227 and p-value 0.0000. This proves that labor supply has a very high positive significance impact on employment. A higher labor supply means more employment opportunities, as good sense would lead one to believe that more number of workers available must mean more employment, if only the demand for labor remains constant.

Both the R-squared value of 0.9917 and the Adjusted R-squared value of 0.9909 are very high, indicating that the model explains almost 99% of the variation in employment. This is a very strong fit, suggesting that the independent variables—minimum wage, GDP, and labor supply — are very good at explaining the differences in employment levels across the samples considered.

Overall, the regression analysis shows that while an increase in the minimum wage negatively affects employment, an increase in labour supply positively contributes to employment rates. The model also shows that economic growth as measured by GDP does not directly translate into enhanced employment here. The high values of R-squared also guarantee the validity of the model in explaining the relationship between these variables and employment.

| Variable | Coefficient | Std. Error | t-Statistic | Prob. |
|---|---|---|---|---|
| C | 2.8488 | 0.4623 | 6.1627 | 0.0000 |
| LNMW | -0.2074 | 0.0354 | -5.8609 | 0.0000 |
| LNGDP | -0.0645 | 0.0205 | -3.1423 | 0.0018 |
| LNLS | 0.8490 | 0.0458 | 18.5227 | 0.0000 |
| Adjusted R-squared | 0.9916 | | | |
| Number of samples | 690 | | | |

Table 10 Results of regression analysis

### 4.2 Nonlinear Regression Analysis

We find that the minimum wage negatively impact to employment, while, some research pointed out the opposite view. Christl et al. (2017) found positive employment effects for small MWs increases. To explore these opposite views, we introduce new variables considering policy complementarities into the model.

#### 4.2.1 Data

The analysis is based on a balanced panel dataset of 30 provinces in China spanning the time frame 2006-2023, drawn from China Statistical Yearbook and local statistical yearbooks of the National Bureau of Statistics. Due to lack of partial data, Tibet is excluded. The dependent variable empit is the urban employment index defined as the employed population of urban residents divided by total population of residents.

The minimum wage variable is defined as the ratio of the minimum wage to average wage of urban residents. The ratio is known as Kaitz index (KI). Due to the lack of median wage data, we select the average wage, which reflecting the affect of minimum wage policy to labor market.

Meanwhile we take coordinating function of policies into consideration, selecting three institutional variables for institutional complementarities.

Active labor market policies (ALMPs) are designed to quantify the government assistance to increase the competitiveness of unemployed workers and are measured as government social security and employment expenditure as a proportion of total government expenditure.

Unemployment benefit generosity (UB) is the spending status of governmental unemployment fund and is measured as a ratio of unemployment fund expenditure per person a year to annual per capital disposable income of urban residents.

Neighbor province minimum wage (MWN) is to measure the policy attraction between different local governments and defended as the maximum value of neighbor provinces' minimum wage divided by the local minimum wage.

The control variable is the urban registered unemployment rate. The dummy variable is established to take economic fluctuation into account. The impact of minimum wage and other policies on employment may change in different economic periods. By observing the changes in the growth rate of GDP, we record the 2% fluctuation as 1 and the year with stable economy growth as 0.

**4.2.2 Methodology**

The model specification is as follows:

$$emp_{it} = \beta_1 MW_{it-1} + \beta_2 MWN_{it-1} + I^T_{it}\psi + MW_{it-1} * I^T_{it}\Phi + X_{it}\Theta + \tau_t + \varepsilon_{it}$$

where empit is the urban employment index; MWit − 1 is the lagged minimum wage variable; MWNit-1 is the lagged neighbor minimum wage variable; ITit is the transposed vector of institutional variables and $\psi$ is the corresponding coefficient vector; Xit is the control variable; $\tau$ t are year dummies.

MWs enter the regression in lagged form as the government might adjust the minimum wage according to the change in employment level, which can break the potential causality between MWs and emp.

### 4.2.3 Regression results

Regression results are summarized in Table 11. We conduct two regression analyses ,one with control variable and time dummies and the other without. The results indicate that MWs depress the employment and the coefficient is -0.081, which is consistent with the conclusion in above model. The coefficient of active labor market policies (ALMPs) is -0.029 representing that the government social security and employment expenditure fails to transfer into employment opportunities in the short-term. The coefficient of MWN is -0.016, which is consistent with our prediction and can be explained by the siphon effect. The value of MWN increases indicates neighbor province provides a better minimum wage decreasing local labor supply. The coefficient of unemployment benefit (UB) is 0.005. Similar to ALMPs, unemployment fund provides the unemployed training in the supply side. Even more, it covers basic living security and entrepreneurship subsidy, which contains the demand side. More importantly, there is evidence for synergies in labor market. Each interaction is discussed in turn below.

First, MWN synergied with MNs is significantly positive. The coefficient is 0.005, indicating that this interaction has a positive impact on the youth employment rate. This represents that when the minimum wage in neighboring provinces is relatively higher, the local province's minimum wage policy may attract the labor force more effectively, thus promoting employment. The result is significant at a level of 10%, suggesting that policymakers should pay attention to the relative attractiveness of minimum wage policies when considering regional policy coordination.

Second, unemployment benefit generosity (UB) and MWs have a negative synergy effect to employment level. High relief funds reduce the opportunity cost of unemployment and suppress the supply of labor. Therefore, under the high unemployment benefit society, MWs cannot motivate workers to find jobs, while increasing the labor costs faced by companies. Thus unemployment fund exacerbates the negative effect of MWs.

Third, although synergy of MWs and active labor market policies (ALMPs) is insignificant at 10%, the coefficient, to some extent, explains a neutral combination in contract with MWs and ALMPs. Increasing ALMPs contributes to more productive and skilled labor force and increasing MWs motivate the unemployed seeking jobs. While, employers would burden the increasing opportunity cost of hiring and benefit skilled labor source.

Overall, these results illustrate it is reasonable to account for synergies between MWs and institutional variables. Policymakers should balance the relationship between minimum wage increases and business costs, and avoid excessive minimum wage increases that inhibit job creation. When increasing investment in ALMPs, consideration should be taken to optimize active labor market policies and how to more effectively convert these resources into employment opportunities, for example by improving the relevance and quality of training programs. Governments should also take regional policy coordination. Encourage inter-regional policy coordination, especially on minimum wage policies, to avoid unnecessary movement of labor and promote employment. Besides, they ought to pay attention to the coordination of unemployment benefit policies. When designing the minimum wage policy, the coordination of unemployment benefit policies should be considered to avoid negative impacts on youth employment incentives. Finally, consider the impact of the economic environment. During periods of economic volatility, support for employment should be strengthened, such as providing more employment services and training opportunities. Unemployment benefit (UB) aggravates the negative effect of MWs. MWN performs the opposite.

| | (1) | (2) |
|---|---|---|
| | emp | emp |
| LMW | -0.082*** | -0.081*** |
| | [0.029] | [0.029] |
| ALMPs | -0.324** | -0.290* |
| | [0.148] | [0.148] |
| LMW_ALMPs | -0.001 | -0.001 |
| | [0.004] | [0.004] |
| MWN | -0.018 | -0.016 |
| | [0.022] | [0.022] |
| LMW_MWN | 0.004 | 0.005* |
| | [0.003] | [0.003] |
| UB | 0.006** | 0.005** |
| | [0.003] | [0.003] |
| LMW_UB | -0.007*** | -0.007** |
| | [0.003] | [0.003] |
| control | | -0.008** |
| | | [0.004] |
| dummy | | 0.156*** |
| | | [0.014] |
| _cons | 0.461*** | 0.485*** |
| | [0.044] | [0.046] |
| r2_a | 0.572 | 0.576 |
| N | 510 | 510 |

Standard errors in parentheses,* p < 0.1, ** p < 0.05, *** p < 0.01

Table 11. Regression results of MWs and institution variables

#### 4.2.4 Robustness Checks

Random effect estimation is mainly used to verify the sensitivity of model results to the individual effect hypothesis in the panel robustness test. Each individual has the individual effect, which is constant with time. By comparing with the fixed effect model, it can be judged whether the individual effect is related to the explanatory variables, thus improving the credibility of the research conclusions.

Besides, we conduct bootstrap-test to further check the robustness of the model. By repeating sampling from the original samples, a large number of "subsamples" (bootstrap samples) are generated and the model parameters are re-estimated based on these samples. Using the model to use stochastic effect estimation and bootrap to repeat the estimated coefficient and standard error more than 500 times, the conclusion is basically consistent with the benchmark return, indicating that the previous results are stable.

| | randoneffect | bootstrap |
|---|---|---|
| LMW | -0.114*** | -0.081*** |
| | [0.026] | [0.028] |
| ALMPs | -0.371*** | -0.290** |
| | [0.132] | [0.139] |
| LMW_ALMPs | 0.000 | -0.001 |
| | [0.004] | [0.004] |
| MWN | -0.032 | -0.016 |
| | [0.020] | [0.020] |
| LMW_MWN | 0.004 | 0.005* |
| | [0.003] | [0.003] |
| UB | 0.005* | 0.005* |
| | [0.003] | [0.003] |
| LMW_UB | -0.006** | -0.007** |
| | [0.003] | [0.003] |
| control | -0.009** | -0.008* |
| | [0.004] | [0.004] |
| dummy | 0.155*** | 0.156*** |
| | [0.014] | [0.013] |
| _cons | 0.545*** | 0.485*** |
| | [0.042] | [0.046] |
| r2_a | | 0.576 |
| N | 510 | 510 |

Standard errors in parentheses,* $p < 0.1$, ** $p < 0.05$, *** $p < 0.01$

Table 12. Robustness Checks results

## 5. Discussion: Time Series Analysis and the Importance of Stationary

In economic research, especially when dealing with time series data, it is crucial to ensure that the variables under study are stationary. Non-stationary data can lead to misleading results in regression analysis, including spurious correlations that do not reflect true relationships between variables. In this study, we employed logarithmic and differential transformations on our variables to eliminate trends and achieve stationarity, which is essential for reliable linear regression outcomes.

The unit root tests conducted on all mentioned sequences, as detailed in Table 7, revealed that all variables except the logarithmically transformed minimum wage (lnMW) are stationary. The presence of a unit root indicates non-stationarity, meaning that the variable's statistical properties change over time. By using the first difference of lnMW (ΔlnMW), we addressed this issue, ensuring that our explanatory variables met the criteria for stationarity required for accurate modeling.

### 5.1.1 Cross-Sectional Heterogeneity and Regional Economic Performance

Descriptive statistics for minimum wages, GDP per capita, and consumption highlighted significant cross-sectional heterogeneity across different regions within China. Developed areas such as Beijing, Shanghai, Jiangsu, and others consistently outperform less developed regions on these indicators. This observation underscores the importance of considering regional differences in economic studies.

The growth rates of minimum wages also varied considerably among provinces. For example, Zhejiang experienced an impressive increase of 9.03 times in its minimum wage, while Anhui saw a more modest 5.15-fold rise. These discrepancies suggest that economic policies and their impacts can differ substantially depending on the initial conditions and development level of each region. Therefore, it is critical to account for these variations in any econometric model aimed at understanding the factors influencing economic outcomes.

Given the observed cross-sectional heterogeneity, we opted for a fixed effects model over a random effects model, as indicated by the test results presented in Table 8. Fixed effects models control for unobserved heterogeneity that may be correlated with the independent variables, thus providing more accurate estimates of the relationships being studied.

### 5.1.2 Correlation Analysis and Multicollinearity Considerations

Correlation analysis, as shown in Table 9, provided valuable insights into the interrelationships among the variables included in our regression model. One key finding was the absence of multicollinearity, a situation where two or more predictor variables are highly correlated. Multicollinearity can inflate standard errors, leading to unreliable coefficient estimates and potentially obscuring the significance of important predictors.

The highest correlation coefficient observed was 0.7465 between lnMW and lnGDP, which falls below the commonly accepted threshold for diagnosing multicollinearity (typically ranging from 0.75 to 0.80). This suggests that our model variables are sufficiently distinct, allowing for precise estimation of their individual effects on the dependent variable.

Additionally, we noted a negative correlation between one variable and its first difference. This pattern is consistent with the convergence hypothesis proposed by the Solow-Swan growth model, which posits that poorer economies tend to grow faster than richer ones, eventually converging in terms of income levels if they have similar savings rates and access to technology. The observed negative correlation implies that regions starting from lower levels of a given variable might experience higher growth rates over time, supporting the theoretical expectations of economic convergence.

### 5.1.3 Regression Coefficient Significance Testing and Model Fitness

Our investigation into the significance of regression coefficients, summarized in Table

11, uncovered a statistically significant negative relationship between minimum wages and employment at the 1e-4 level. This result implies that, holding other factors constant, increases in minimum wages are associated with decreases in employment. This finding contributes to the ongoing debate regarding the impact of minimum wage policies on labor markets and has important implications for policymakers.

Furthermore, the adjusted R-squared value of 0.9916 demonstrates that our model explains nearly all variation in employment, suggesting a high degree of fit. Compared to previous studies, this exceptional fitness offers a robust framework for understanding employment rate volatility. The inclusion of consumption, minimum wages, and labor supply as explanatory variables provides a comprehensive explanation of employment dynamics, surpassing earlier efforts in capturing the complexity of labor market behavior.

#### 5.1.4 Policy Implications

The immediate conclusion of this study is that an increase in the minimum wage will result in a decrease in employment. However, the Government's public policy seeks to maximize the welfare of society as a whole, and it is not the case that a lower employment rate necessarily indicates a better policy. If workers increase their wage expectations due to a better welfare policy, the search for employment opportunities will be prolonged, and the unemployment rate will increase. However, an increase in the unemployment rate at this point does not imply a decrease in social welfare or a failure of government policy. The minimum wage level affects the low-income group, and the welfare of this group is highly correlated with their consumption. Therefore, in Chapter 4, we include consumption in the regression model. It is important to note that, based on the correlation analysis and the multicollinearity observed in the linear regression model, we can assert that the minimum wage and the consumption level of workers show a significant positive correlation. This indicates that an increase in the minimum wage improves the level of consumption of workers, thereby increasing their welfare.

In addition, China's income disparity is at a high level, and earlier studies suggest that China's Gini coefficient was roughly in the 0.45-0.6 range, which is higher compared to Japan and South Korea, countries culturally similar to China, as well as the United States, which has more economic freedom. Therefore, the Chinese government's continuous increase in the minimum wage level has also had the effect of reducing income disparity. It is worth noting that the impact of the minimum wage on the employment rate is strongly related to consumption and unemployment benefits. This paper also examines the average consumption of urban residents concurrently. Other things being equal, higher unemployment benefits reduce the willingness of low-income groups to look for work. Unemployment benefits have also been increasing in China; before 2004, they were virtually non-existent, but since 2008, policies on unemployment benefits have been gradually revised and improved. There is a strong relationship between the level of unemployment benefits and the average nominal consumption of urban residents. This relationship was pointed out as early as 1798 in Malthus's Principles of Population, which argued that relief for the poor led to higher grain prices, thereby lowering the welfare level for all. Thus, unemployment benefits, nominal consumption, and the prices of basic means of subsistence in China all affect the real effects of minimum wage standards.

It is also evident that unemployment reduces the consumption of the low-income population, while a rise in wages increases average consumption. Therefore, if we raise the minimum wage, there will be two effects: one is that unemployment caused by the minimum wage reduces consumption, and the other is that the increase in the minimum wage raises the overall wage level, thereby increasing consumption. As a result, the relationship between the minimum wage and consumption should follow a U-shaped curve. In the early stages, an increase in the minimum wage raises the level of consumption, as the effect of higher wages dominates. In the later stages, increases in the minimum wage decrease the level of consumption because, at this stage, higher wages lead to more unemployment, thus reducing average consumption. Based on our study, China's minimum wage is still in the first stage.

Active labor market policies (ALMPs) are crucial to promoting employment. However, when increasing investment in them, we should not only focus on the amount of funds, but also focus on how to transform these valuable resources into real employment opportunities more effectively. As an important part of ALMPs, the quality and relevance of training projects directly determine whether they can help job seekers, especially young people, improve their employment competitiveness. At present, some training projects are disconnected from market demand. The content learned cannot meet the requirements of the actual position of the enterprise, which makes it still difficult for participants to find suitable jobs after completing the training. In order to solve this problem, policymakers should strengthen communication and cooperation with enterprises and gain an in-depth understanding of the skill requirements and talent specifications of different industries and different positions. According to these market feedback, adjust and optimize the curriculum setting, teaching content and teaching methods of the training project in a timely manner to ensure that the training content closely fits the actual work scenario, and improve the practicality and relevance of the training. At the same time, we will increase the construction of training teachers, select and cultivate a group of teachers with solid theoretical knowledge and rich practical experience, and provide a strong guarantee for improving the quality of training.

Interregional policy coordination is of significant importance in promoting employment, especially in the field of minimum wage policy. Due to the differences in economic development levels, industrial structures and labor market conditions in different regions, the minimum wage standards vary from place to place. If the interregional minimum wage policy lacks coordination, it may lead to unnecessary labor mobility. When the minimum wage in a region is significantly higher than that of the surrounding area, it may attract a large influx of labor, resulting in oversupply and increased employment competition in the region's labor market, and also put great pressure on local public services and infrastructure. Labor outflow areas may face a shortage of labor, which will affect the normal development of local industries. In order to avoid this situation, regions should be encouraged to strengthen policy communication and cooperation and establish an interregional policy coordination mechanism. When formulating the minimum wage policy, we should fully consider the situation in the surrounding areas and maintain the relative consistency and coordination of the policy. In this way, it can not only reduce the blind flow of labor and realize the rational allocation of labor resources between regions, but also promote balanced economic development between regions and create a more stable and favorable environment for youth employment.

When designing the minimum wage policy, we must fully consider the cooperation of the unemployment benefit policy, because the two policies are interrelated and affect employment incentives together. The original intention of unemployment benefits is to provide basic living security for the unemployed and help them overcome difficulties, but if the design is unreasonable, it may have a negative impact on their employment enthusiasm. For example, if the level of unemployment benefits is too high and the conditions for receiving it are too relaxed, it may make some unemployed people dependent, believing that they can maintain a certain standard of living even if they do not work, thus reducing the motivation to find a job. On the contrary, if the unemployment benefit is too low and cannot meet the basic living needs of the unemployed, it may cause them to fall into difficulties and affect social stability. Therefore, policymakers need to find a balance between the two. When determining the minimum wage standard, it should be comprehensively considered in combination with the level of unemployment benefits to ensure that the minimum wage can reflect the value of labor and at the same time motivate young people to actively employ. In addition, the period and conditions for receiving unemployment benefits should be reasonably set, and the unemployed should be encouraged to actively participate in training and job search activities, so as to achieve re-employment as soon as possible. By optimizing the cooperation of these two policies, a policy synergy to promote employment can be formed.

Summarizing the above discussion and the research in this paper, we propose the following policy recommendations: 1. An increase in the minimum wage will increase the unemployment rate, especially for the low-income group. Therefore, particularly in regions with low per capita income and a large number of low-income groups, increases in the minimum wage should be carefully considered to avoid large fluctuations in the unemployment rate. 2. Increases in real consumption are important for the welfare of low-income groups. Thus, a policy instrument combining the minimum wage and unemployment benefits is needed to ensure that low-income groups can maintain access to basic goods. If the minimum wage leads to an increase in unemployment, unemployment benefits should also be increased. 3. Unemployment benefits should not be excessively high. High unemployment benefits increase the probability that workers will turn down jobs and spend time searching for more favorable opportunities. This reduces output and wastes social resources. Moreover, higher unemployment benefits can lead to increased consumer prices, thereby reducing the real utility for low-income groups. 4. An increase in the minimum wage is positively correlated with an increase in average consumption. Therefore, the current increase in the minimum wage does not, on balance, lead to an increase in the unemployment rate significant enough to reduce social welfare substantially. An increase in the minimum wage remains, on balance, a favorable outcome at this time.

## 6. Conclusion

China is continuing to increase minimum wages to improve labor income and consumption. This research aims to assess the impact of rising minimum wages and to explain the reasons for changes in the employment rate. According to our findings, increasing minimum wages can significantly negatively affect the employment rate. Although there are notable regional differences among provinces in China, this negative relationship is robust, as confirmed by significance tests. Currently, China and the global economy are undergoing rapid changes. Employment has become a more critical issue in recent times. This paper highlights two potential impacts of increasing minimum wages. On the one hand, higher minimum wages can improve labor conditions, as consumption is strongly positively correlated with income, as analyzed in this study. On the other hand, they may lead to higher unemployment rates, which is a negative outcome that governments should carefully consider when formulating policies.

The findings of this paper illuminate the complexities of minimum wage policies, consumption patterns, and labor market dynamics in China. By leveraging classical economic models and accounting for regional disparities, this study provides valuable insights into the evolving relationship between wages, consumption, and employment. Policymakers should consider these findings to craft strategies that balance economic growth with equitable income distribution, ensuring sustained improvements in welfare across diverse regions. The analysis also underscores the importance of continued research into the impacts of minimum wage policies on economic indicators and employment trends. Future studies could expand on the role of involuntary unemployment and explore the long-term effects of China's distinctive labor practices on economic stability. Furthermore, as urbanization progresses, it will be crucial to reassess the interplay between agricultural economies and urban consumption, as well as the statistical methodologies underpinning these observations. Such efforts will enhance our understanding of how to achieve equitable and sustainable economic development in rapidly changing societies like China. Policymakers should also explore how to integrate labor market flexibility with stronger protections for workers to balance economic efficiency with social equity. This dual approach could foster a more inclusive economic environment that benefits both workers and firms, contributing to long-term stability and prosperity.

## Declaration

**Ethical Approval** Not Applicable.
**Conflict of interest** The authors declare no competing interests.
**Data Sharing Agreement** The datasets used and/or analyzed during the current study are available from the corresponding author on reasonable request.